# Relativity in Classical Mechanics: Momentum, Energy and the Third Law

Roberto Assumpção, PUC-Minas, Poços de Caldas- MG  37701-355, Brasil
*assumpcao@pucpcaldas.br*

***Abstract*** *– Most of the logical objections against the classical laws of motion, as they are usually presented in textbooks, centre on the fact that defining force in terms of mass and acceleration, the first two laws are mere assertions of concepts to be introduced in the theory; conversely, the third law expresses the experimental fact that the ratio of masses is inversely proportional to the ratio of accelerations, but it is known to fail when the interacting bodies are rapidly accelerated or far apart, leading to objections at the research level, particularly when electromagnetic phenomena is present. Following a specification of the coordinate system with respect to which velocities and accelerations are to be measured, relative to a fixed spacetime point, this contribution argues that the limitation of the third law is removed; as a consequence, Energy and Momentum relations are given an alternative formulation, extending their fundamental aspects and terms to the relativistic level. Most important, the presented alternative relations seem to preserve exactly the same form of the concepts as originally used by Newton in the Principia.*

## Introduction

The classical laws of motion [1,3] follow from the analysis of a "free–fall" experiment (isolated bodies) whose result, the ratio of masses inversely proportional to the ratio of accelerations, can be generalised to include "any" pair of bodies:

$$\frac{m}{M} = \frac{\dot{V}}{\dot{v}} \quad (1)$$

where V refers to a body of mass M and *v* to a body of mass *m*, the dot meaning the usual time differentiation.

Actually, accelerations are the experimental data, so that this should be written as:

$$\frac{m}{M} = \frac{\dot{V}_{exp}}{\dot{v}_{exp}} \quad (1.a)$$

where the subscript "*exp*" explicitly indicates that the right hand ratio is an experimental or measured quantity.

Now, instead of a discussion concerned with the elements ( force, momentum, kinetic energy) that led to the construction of the theory, the emphasis here will be on the fact that the framework of classical mechanics is based on experimental data; this points to a comprehensive method of acceleration measurements in spacetime.

## Spacetime Measurements

According to classical mechanics, objects flow in spacetime, in a sense that coordinates can be specified as functions of time, such as x = x(t); assuming the meaning of x(t) is known, velocities and accelerations are taken as usual:

$$v = \frac{\Delta x}{\Delta t} = \frac{dx}{dt} = \dot{x} \quad ; \quad a = \frac{\Delta v}{\Delta t} = \frac{dv}{dt} = \dot{v} = \ddot{x} \quad (2)$$

However, from the experimental point of view, if objects flow in time, the experimental values of velocities and accelerations are obtained via time measurements, so that relations (2) should be written as:

$$v_{exp} = \frac{\Delta x}{\Delta t_{exp}} \quad ; \quad a_{exp} = \frac{\Delta v}{\Delta t_{exp}} \quad (2.a)$$

In relations (2.a) the time derivatives are absent due to the fact that they are theoretical functions whereas the Δ symbolises only experimental data. According to this relation, velocity measurements imply detection of two events $\Delta t = t_2 - t_1$ taking place at distinct places, $\Delta x = x_2 - x_1$, as shown bellow:

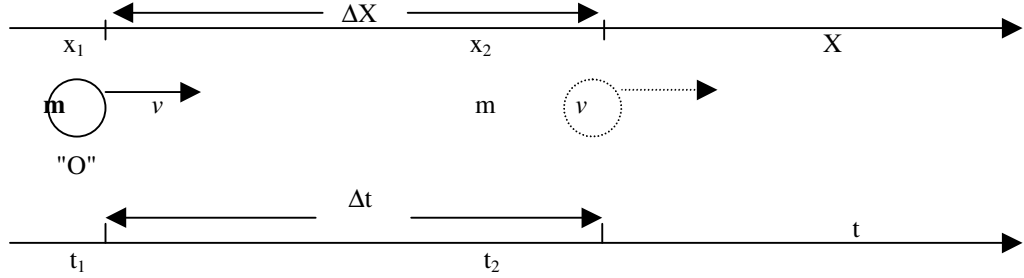

Figure 1
Velocity measurement employing space and time detection. Observer "O" is assumed at ($x_1$, $t_1$), the position of the first measurement.

Now, the distance $\Delta x = x_2 - x_1$ can be determined *a priori* or *a posteriori*; however, the temporal location must be determined "on time"; this states that the time measured ($\Delta t_{exp}$) by the observer at $t_1$ is the elapsed time ($\Delta t = t_2 - t_1$) of the two events plus the time required to transport the event taking place at ($x_2$, $t_2$) to ($x_1$, $t_1$):

$$\Delta t_{exp} = \Delta t + \Delta t_i \quad (3)$$

where $\Delta t_i$ is the time required to transport the ($x_2$, $t_2$) event to ($x_1$, $t_1$); from (2.a) it follows that

$$\frac{\Delta x}{v_{exp}} = \frac{\Delta x}{v} + \frac{\Delta x}{v_i} \quad (4)$$

where $v_i$ is the informational velocity or the speed of the signal connecting ($x_2$, $t_2$) to ($x_1$, $t_1$) and $v_{exp}$ the experimental value of $v$. Eq. (4) can be written as

$$\frac{1}{v_{exp}} = \frac{1}{v} + \frac{1}{v_i} \quad (5)$$

It then follows that

$$v_{exp} = \frac{v}{1 + \frac{v}{v_i}} \quad (6)$$

implying that velocity measurements underestimates the real value, so that the classical assumption that, in principle, it is possible to devise instruments to measure this quantity with as small an error as we please holds only if the speed $v_i$ is infinity. Assuming the speed of light in vacuum – c as a limiting experimental speed, it can be shown [2] that:

$$v = \frac{v_{exp}}{\left(1 - \frac{v_{exp}}{2c}\right)} \quad (7)$$

where $v$, the "true" or theoretical value of the speed of the object is written in terms of experimental quantities. Time equation (3) is therefore

$$\Delta t = \left(1 - \frac{v_{exp}}{2c}\right) \Delta t_{exp} \quad (8)$$

The pair of equations (7) and (8) gives a practical way of measuring time, velocities and consequently accelerations; the last can be obtained by dividing (7) by (8)

$$\dot{v} \cong \frac{\dot{v}_{exp}}{\left(1 - \frac{v_{exp}}{2c}\right)^2} \quad (9)$$

where the left term is the "true" or theoretical value of the acceleration written in terms of experimental values; please note that $v_{exp}$ is an experimental quantity, so the dot over it is just representative. Now, this result can be employed to analyse the free-fall .

**Momentum**

From (1.a), it follows that we have to balance the accelerations of the two bodies:

$$\dot{v}_{exp} = \dot{v}\left(1 - \frac{v_{exp}}{2c}\right)^2 \quad \text{acceleration of body } m$$

$$\dot{V}_{exp} = \dot{V}\left(1 - \frac{V_{exp}}{2c}\right)^2 \quad \text{acceleration of body } M$$

Substituting in (1.a), the ratio of masses becomes

$$\frac{m}{M} = \frac{\dot{V}\left(1 - \frac{V_{exp}}{2c}\right)^2}{\dot{v}\left(1 - \frac{v_{exp}}{2c}\right)^2}$$

Rearrangement of the right hand gives

$$\frac{\left(1 - \frac{V_{exp}}{2c}\right)^2}{\left(1 - \frac{v_{exp}}{2c}\right)^2} = \frac{(2c - V_{exp})^2}{(2c - v_{exp})^2}$$

so that

$$\frac{m}{M} = \frac{\dot{V}(2c - V_{exp})^2}{\dot{v}(2c - v_{exp})^2}$$

or

$$\frac{M\dot{V}}{m\dot{v}} = \frac{(2c - v_{exp})^2}{(2c - V_{exp})^2}$$

where the right hand contains only experimental quantities; it then follows that

$$\frac{\sqrt{M\dot{V}}}{\sqrt{m\dot{v}}} = \frac{2c - v_{exp}}{2c - V_{exp}} \quad (10)$$

Defining an "instrumental function" $f = f(v_{exp}, V_{exp}, c)$ as

$$f \equiv \frac{2c - v_{exp}}{2c - V_{exp}} \quad (11)$$

we note* that assuming the limiting experimental speed as that of light c, $f$ can be set equal to 1 without so much error. Therefore equation (10) can be written as,

$$m\dot{v} \cong \sqrt{M\dot{V}m\dot{v}} \quad (10.a)$$

The left hand is a force; this impressed force is sensed by body $m$ ; following the original newtonian quotation [3], the change of motion ( usually taken as the rate of change of momentum – $P$ ) is proportional to the impressed force, therefore

$$\dot{P} = \frac{\Delta P}{\Delta t_{int}} \approx \sqrt{M\dot{V}m\dot{v}} \quad (11)$$

where $\Delta t_{int}$ is the time of interaction of the two bodies $m$ , M . Integration of equation (11) is carried over this period of time, so that

$$P = \iint \dot{P}\Delta t_{int} \approx \iint \sqrt{M\dot{V}m\dot{v}}\Delta t_{int}$$

By noting that $\Delta t_{int}$ is the rate of change of the two body system – [ $m$ , M ], it is taken as a medium value of the "time of flight" of the individual bodies

$$\Delta t_i = \sqrt{dt_V \cdot dt_v} \qquad (12)$$

where $dt_V$ is the proper time of M and $dt_v$ the proper time of $m$. Therefore,

$$P \cong \iint \sqrt{Mm}\sqrt{\dot{V}dt_V}\sqrt{\dot{v}dt_v}$$

adopting the definition of acceleration as

$$\dot{V} \equiv \frac{d_V}{dt_V}$$

we can write

$$\sqrt{\dot{V}dt_V} = \sqrt{d_V}$$

so that the double original temporal integration can be substituted by individual ones carried on velocities

$$P \cong \sqrt{Mm}\int_V\int_v \sqrt{dv}\sqrt{dV}$$

employing the summation rule for square roots, a rough tough exact result is

$$P \cong \sqrt{MVmv} \qquad (13)$$

implying that momentum is a "dual" quantity representing the [ $m$ , M ] system; not individual particles. *The mutual actions of two bodies upon each other* produces momentum.

The result resembles an impulse-type force that depends only on velocities, though a spacetime representation can be sketched. In one dimension this is equivalent to two component equations, which for similar masses (and velocities) reads ( Figure 2):

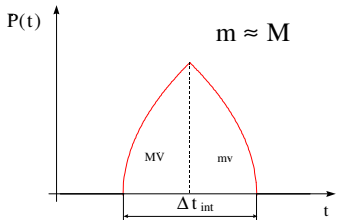

Figure 2
Momentum representation in spacetime; the double square root is plotted for each component, considered as bodies of similar magnitude ( m$v$ ~ MV )

The symbolic vectors V and $v$, according to the definitions recorded here, can also be expressed as experimental quantities; in this sense, the "Third Law" admits a distinct interpretation.

**The Third Law**

The failure of the third law [4] is commonly [1] associated to electromagnetic phenomena, but in general it fails for any forces which propagate from one particle to another with finite velocities; this failure can be associated to the classical assumption that the behaviour of measuring instruments is unaffected by their state of motion, leading to the foundations of the theory of relativity. But one can express the momentum result in terms of experimental quantities; in this manner, taking eq. 11, the impressed force becomes:

$$\dot{P} = \sqrt{M\dot{V}m\dot{v}} = \sqrt{Mm}\sqrt{\frac{\dot{V}_{exp}}{\sqrt{\left(1-\frac{V_{exp}}{2c}\right)^2}}}\sqrt{\frac{\dot{v}_{exp}}{\sqrt{\left(1-\frac{v_{exp}}{2c}\right)^2}}}$$

where the right hand contains only experimental data.
Noting that the impressed force is sensed by body $m$, this can be rewritten as

$$F_v = \frac{1}{\left(1-\frac{V_{exp}}{2c}\right)}\frac{1}{\left(1-\frac{v_{exp}}{2c}\right)}\sqrt{Mm\dot{V}_{exp}\dot{v}_{exp}} \qquad (14)$$

where $F_v$ refers to body $m$.

Equation (14) gives a way of specifying forces which propagate from one body to another with finite velocities; it is an active picture of the original newtonian formulation that incorporates velocity-dependants non-Lorentzian factors ( the pre-square root terms); nevertheless, these are active only as measured velocities approach the speed of light.

It is difficult to oppose the distinguished authority consecrated by centuries of usage; however, specification of the coordinate system with respect to which accelerations are to be measured ( relative measurement determined by the signal carrier, figure 1 ) sets the usual interpretation of the third law, action equals reaction, as a limiting static situation. We now consider energy relations.

**Energy**

The concept of energy and its *qualifying adjectives* changed historically and also from discipline to discipline [4,5] ; perhaps the less elegant tough more physically sounded is the (kinetic) correspondence to work. In this sense, it is associated to the spatial rate of change of the impressed force:

$$E \equiv \int dx F_v = \int dx \sqrt{M\dot{V} m\dot{v}} \quad (15)$$

from the definition of velocities and accelerations,

$$V \equiv \frac{dx}{dt_V} \;;\; \dot{V} \equiv \frac{dv}{dt_V}$$

and noting that
$dx \equiv V dt_V \;;\; dx \equiv v dt_v$ so that $dx^2 \equiv V dt_V v dt_v$
(15) can be written as

$$E \equiv \iint \sqrt{M\dot{V} m\dot{v}} \sqrt{V dt_V} \sqrt{v dt_v}$$

so that

$$E \approx \sqrt{Mm} \sqrt{\iint V dV v dv}$$

or

$$E \approx \frac{1}{2} \sqrt{Mm} \, vV \quad (16)$$

Again the result applies to the pair [ M, m] ; expressing the last equation in terms of experimental quantities, we obtain :

$$E \approx \frac{1}{2} \frac{1}{\left(1 - \frac{V_{exp}}{2c}\right)} \frac{1}{\left(1 - \frac{v_{exp}}{2c}\right)} \sqrt{Mm} \, v_{exp} V_{exp} \quad (17)$$

These equations are analogous to the conventional ones when the masses and velocities of the interacting bodies are exactly equal; conceptually, the procedure to go over from the usual classical treatment of the motion to the one recorded here is straightforward: rather than assuming the equality between theoretical and experimental quantities, such as *v* and *v_{exp}* , one considers that the last can be obtained only via a third element, the informational velocity $v_i$. On the other hand, it was the motivation for the analysis of experimental data which generate the symbolic or theoretical terms, such as E in eq. (17)

**Conclusions**

Despite the conventional status of Newton's laws, assumed to hold only in a limited range, particularly the third one, the recorded results suggest that their fundamental aspects and terms can reach the level of relativistic mechanics, provided a different scheme of the measurement problem is settled. Assuming the basis of Classical Mechanics as the experimental fact that the ratio of masses is inversely proportional to the ratio of accelerations <u>and</u> specifying the coordinate system with respect to which velocities and accelerations are to be measured, a clear distinction between experimental and theoretical entities is achieved. This leaves the elements usually employed in the construction of the theory, such as force, momentum, kinetic energy, etc.., to a second plane, emphasising the fact that the framework of classical mechanics is based on experimental data. It is hoped that this approach is adequate to Experimental Physics and Engineering courses in addition to the importance of incorporating advanced topics and integrating the different branches of mechanics.

**Notes :**

1. The experimental or instrumental function $f = f(v_{exp}, V_{exp}, c)$ defined as

$$f \equiv \frac{2c - v_{exp}}{2c - V_{exp}}$$

varies between ½ and 2 due to the fact that experimental speeds, $v_{exp}$ and $V_{exp}$ are limited by c. Thus the medium (geometrical) value is 1, corresponding to classical (<<c) velocities . A similar function appears in the analysis of a "free-fall" experiment [7] .

2. *Change of motion* differs from the common statement *rate of change of momentum* , though the symbolic "*P*" is used for the last; note however that eq. (13) represents *the mutual action of two bodies upon each other*, a statement appearing in the third law [ 3,6].